\documentclass{iau} 
\usepackage{graphicx}

\title[Supermassive Black Hole Scaling Relations] 
{Coevolution (Or Not) of Supermassive \\ Black Holes and Host Galaxies: \\
Black Hole Scaling Relations Are Not Biased by Selection Effects}

\author[John Kormendy]  
{John Kormendy$^{1,2,3}$}

\affiliation{$^1$Max Planck Institute for Extraterrestrial Physics, \\
    Giessenbachstrasse, D-85748 Garching by Munich, Germany \\[\affilskip]
$^2$Munich University Observatory, Scheinerstrasse 1, D-81679 Munich, Germany \\ [\affilskip]
$^3$Department of Astronomy, University of Texas at Austin \\
    2515 Speedway, Mail Stop C1400, Austin, TX 78712-1205, USA \\ 
    email: {\tt kormendy@astro.as.utexas.edu}}

\pubyear{2019}
\volume{353}  
\jname{Galactic Dynamics in the Era of Large Surveys}
\editors{M.~Valluri, \& J.~A.~Sellwood, eds.}
\begin{document}

\maketitle

\begin{abstract}
The oral version of this paper summarized Kormendy \& Ho 2013, ARA\&A, 51, 511.
However, earlier speakers at this Symposium worried that selection effects
bias the derivation of black hole scaling relations.  I therefore added -- and this
proceedings paper emphasizes -- a discussion of why we can be confident that selection
effects do not bias the observed correlations between BH mass $M_\bullet$ and the
luminosity, stellar mass, and velocity dispersion of host ellipticals and classical
bulges.  These are the only galaxy components that show tight BH-host correlations.
The scatter plots of $M_\bullet$ with host properties for pseudobulges and disks are
upper envelopes of scatter that does extend to lower BH masses.  BH correlations are
most consistent with a picture in which BHs coevolve only with classical bulges and
ellipticals.  Four physical regimes of coevolution (or not) are suggested by Kormendy
\& Ho 2013 and are summarized here.
\end{abstract}

\firstsection 

\centerline{\null} \vskip -28pt \centerline{\null}

\section{Introduction}

\pretolerance=15000  \tolerance=15000
\def\ts{\thinspace}

      This is two papers in one, as reflected in the title:

      My original aim was to summarize the Kormendy \& Ho (2013) review of supermassive black hole (BH) 
mass measurements and their use to investigate whether host galaxies are influenced by radiative or 
kinetic feedback while BHs grow by accretion as active galactic nuclei (AGNs).

      However, other speakers at this conference (e.{\ts}g.,~Bureau 2019) echo published papers
that cast doubts on measurements of BH mass $M_\bullet$ scaling relations.  They claim that we are biased
toward high $M_\bullet$ because BH spheres of gravitational influence are small, so only the
highest-$M_\bullet$ BHs are preferentially discovered. I therefore added a demonstration that the BHs which
satisfy tight $M_\bullet$\ts--{\ts}host galaxy correlations{\ts}--{\ts}and {\it only} those BHs{\ts}--{\ts}are 
discovered in classical bulges and ellipticals with no significant bias in favor of special, 
compact galaxies with respect to fair samples that include more diffuse objects.  I was limited in how much detail 
I could include in real time.  Here, Section 3 enlarges on the evidence that derived $M_\bullet$\ts--{\ts}host 
galaxy correlations are not biased by selection effects. 

      A point of casting doubt can be to introduce a new technique that comes~to~the~rescue.  Bureau (2019) 
emphasizes that molecular gas kinematic measurements with ALMA, the Atacama Large Millimeter Array, can have
higher spatial resolution and certainly have different and better controlled measurement systematics than does 
optical spectroscopy.  Also, measurements of nuclear megamaser disks are immune from optical atmospheric blurring. 
They can find smaller BHs farther away.  In Section 4, I add to published BH scaling relations the new BH 
detections{\ts}--{\ts}most of them from molecular gas kinematics{\ts}--{\ts}reported at this meeting.  
I show that the molecular disk mass measurements define the same BH correlations as do stellar and ionized 
gas dynamic measurements.  This is also true for BH masses derived from maser disk dynamics.
 
\vskip -26pt \null

\section{M{\lower 2pt \hbox{$\bullet$}} Measurements. I. Halo Dark Matter and Broad Emission Lines}
 
      Kormendy \& Ho (2013) review $M_\bullet$ measurements.  Two techniques require ``tweaks'':

      The state of the art for stellar dynamical $M_\bullet$ measurement is Schwarzschild (1979) orbit superposition 
modeling.  This now includes the superposition of tens of thousands of stellar orbital density distributions and 
the fitting of line-of-sight velocity distributions from two-dimensional spectroscopy.  Another improvement is the
addition of halo dark matter (DM) to dynamical models 
(Gebhart \& Thomas 2009;
Schulze \& Gebhardt~2011;
Gebhardt et al.~2011;
Rusli et al.~2013).
Adding DM at large radii causes us to decrease the stellar mass-to-light ratio. Because we assume that
mass-to-light ratio is independent of radius, we decrease the mass-to-light ratio at small radii, too.  Then we 
have to increase $M_\bullet$ in order to continue to explain high velocities there.  This proves to make a bigger 
difference if the black hole sphere of influence is poorly resolved (Schulze \& Gebhardt 2011; Rusli et al.~2013).
It also tends to be more important for core-nonrotating-boxy ellipticals (see Kormendy et al.~2009 for a summary
of the division into core and coreless ellipticals).  These effects and the improvements in orbit sampling 
increase our estimates of $M_\bullet$, in many cases by factors of $\sim$\ts2.
In particular, the BH in M{\ts}87 is found to have a mass of $M_\bullet = (6.2 \pm 0.4) \times 10^9$ 
$M_\odot$ (Gebhardt \& Thomas 2009; Gebhardt et al. 2011).

      This is 1.7 times larger than $M_\bullet = (3.6 \pm 1.0) \times 10^9$~$M_\odot$ given by Hubble Space 
Telescope (HST) spectroscopy of the emission-line rotation curve $V(r)$ 
(Macchetto et al. 1997).  Similarly, the stellar-dynamic $M_\bullet$ is larger than the gas-dynamic estimate in NGC\ts3998. 
In both cases and for several other galaxies with emission-line HST $M_\bullet$ measurements, the authors noted 
that the emission line widths are comparable to the velocity amplitudes.  These widths were ignored in the analysis, 
based on the assumption that gas clouds may have high internal velocity dispersions but may rotate around the 
center at lower~$V(r)$.  This is dangerous:~Nobody makes stellar dynamical models ignoring~velocity~dispersions.  
Now, we have independent confirmation of the M{\ts}87 BH mass from the Event Horizon Telscope collaboration 
(2019), $M_\bullet$\ts=\ts(6.5\ts$\pm$\ts0.2\ts$\pm$0.7)\ts$\times$\ts$10^9$\ts$M_\odot$. 
{\it Kormendy \& Ho 2013 omit from BH correlations nine $M_\bullet$ measurements that they conclude
are underestimated because broad emission-line widths were ignored.}  Caution:~these underestimated masses 
are still included\ts--{\ts}and produce a bias\ts--{\ts}in many published derivations of BH correlations.

\vskip -26pt \null

\section{M{\lower 2pt \hbox{$\bullet$}} Measurements. II. Are Sample Selection Effects Important?}
 
      The answer is ``no'' for the classical bulges and ellipticals that satisfy tight 
$M_\bullet$\ts--{\ts}host correlations but ``yes'' for pseudobulges and disks that do not satisfy such correlations.

      At this meeting, Bureau (2019) and others echo van den Bosch et al.~(2015) who conclude that BHs are discovered
only in ``the densest of galaxies, which [are] not representative of the galaxy population at large.  This is evident 
from the distribution of host galaxy properties \dots~shown in [their] Figure 8.  It is striking how the host 
galaxies trace out a very narrow locus in this parameter space.  This is most obvious in the luminosity--size panel, 
where they lie along a narrow line, sampling preferentially the densest galaxies.  Note that the black hole 
host galaxies are typically denser than the average (early-type) galaxies'' (see also van den Bosch 2016).

\vfill\eject

      Figure\ts1 shows projections of the fundamental plane correlations for  ellipticals and classical 
bulges; $r_e$ vs.~$M_V$ is the bottom panel.  BH detections sample the complete range of parameters 
for bulges and ellipticals.  We do not preferentially find~BHs in the most compact galaxies.  Contrast 
unrelated spheroidal galaxies:~they are more diffuse.  In Figure 1, the scatter for the BH host galaxies 
is actually smaller than the scatter for the

\begin{figure}[b]
\vspace*{-2.0 cm}
\begin{center}
\includegraphics[width=4.5in]{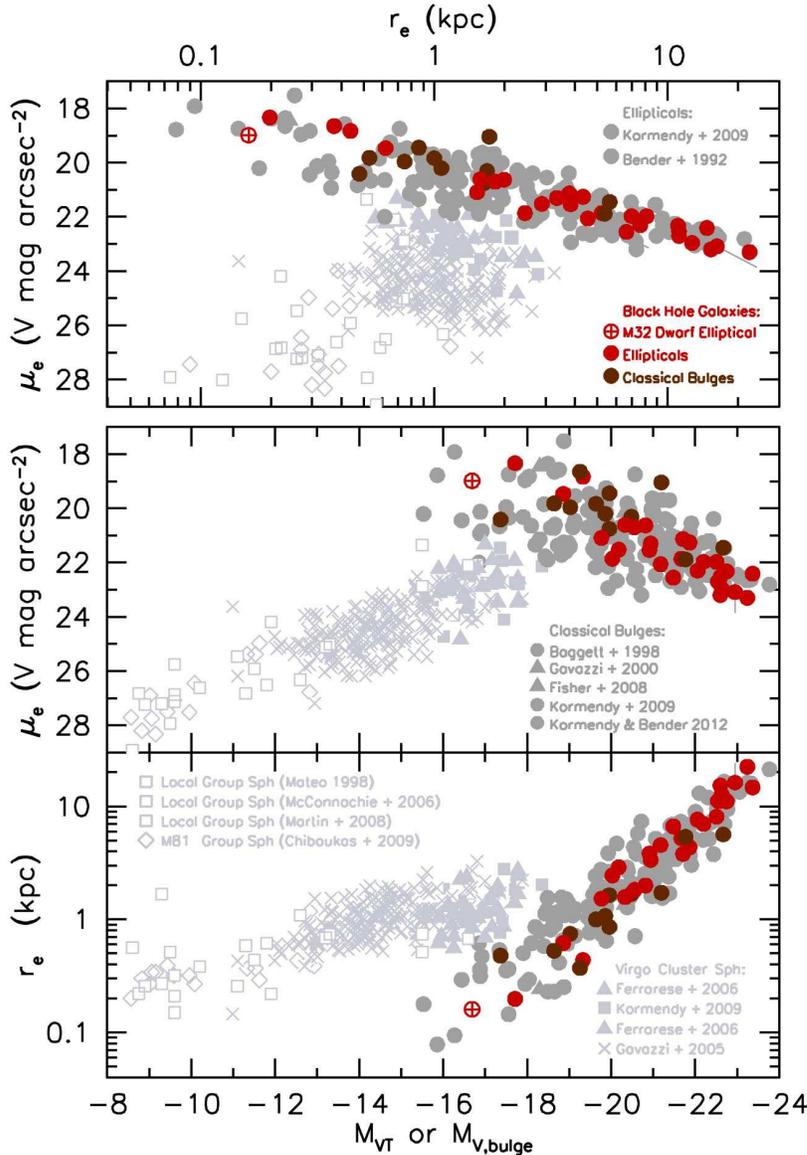} 
\end{center}
\vspace*{-0.3 cm}
\caption{Parameter correlations for elliptical galaxies and classical bulges of disk galaxies (dark gray)
and for spheroidal galaxies (lighter gray).  The bottom panels show the effective radius~$r_e$ that 
contains one-half of the $V$-band light and the effective surface brightness $\mu_e$ at $r_e$ as functions 
of the total $V$-band absolute magnitude of the galaxy component.  The top panel shows the
Kormendy (1987) relation, $\mu_e$ versus $r_e$; this projection shows the fundamental plane nearly edge-on 
and has especially small scatter.  Sources are given in the keys.  Galaxies in which supermassive black holes 
are detected via spatially resolved stellar or gas dynamics are encoded in dark red for ellipticals and 
dark brown for bulges.  This figure is adapted from Figure 16 of Kormendy \& Bender (2012), who provide 
the references in the keys for Sph galaxies.}
\label{fig2}
\end{figure}

\newpage

\noindent galaxies that define the fundamental plane, both toward compactness (small $r_e$ and bright $\mu_e$) and
toward diffuseness (large $r_e$ and faint $\mu_e$).  The reason is that the BH hosts tend to be relatively
nearby and are studied more accurately (e.{\ts}g., in Kormendy et al.~2009, where many sources of photometry are
combined) than the more heterogeneous data on more distant ellipticals and bulges.  The fundamental plane
is well known to have small intrinsic scatter perpendicular to the plane (e.{\ts}g., J\o rgensen et al.~1996).  
That's why BH hosts ``trace out a very narrow locus in [$r_e$\ts--\ts$M_V$] space''
(van den Bosch's words).

      Shankar et al.~(2016) also worry about $M_\bullet$ bias:~``We confirm that 
the majority of black hole hosts have significantly higher velocity dispersions $\sigma$ than local galaxies
of similar stellar mass.  We use Monte Carlo simulations to illustrate the effect on black hole scaling 
relations if this bias arises from the requirement that the black hole sphere of influence must be resolved 
to measure black hole masses with spatially resolved kinematics.  We find that this selection effect artificially
increases the normalization of the $M_\bullet$\ts--\ts$\sigma$ relation by a factor of at least $\sim$3; the 
bias for the $M_\bullet$\ts--\ts$M_{\rm bulge}$ relation is even larger.''

      Figure 2 shows the Faber-Jackson (1976) correlation between $\sigma$ and absolute magnitude from 
Kormendy \& Bender (2013).  I use this because the galaxies are relatively nearby and well studied;
e.{\ts}g., HST was used to look for cores.  BH hosts tend to be high in luminosity; we do not search for BHs 
in distant, tiny ellipticals for which we know we don't have sufficient spatial resolution.  But the $\sigma$ 
{\it scatter} is fairly sampled at least at $M_V > -22$.  The BH host at $M_V \simeq -18$ and $\sigma \simeq 170$ 
km s$^{-1}$ is NGC 4486B; this is one of the ``BH monsters'' that lie far above the BH correlations in Figure 3.
At $M_V < -22$, the core ellipticals that are known to be BH hosts do have slightly higher $\sigma$ than the
\phantom{000000000000}

\begin{figure}[b]
\vspace*{-2.2 cm}
\begin{center} 
 \includegraphics[width=5.2in]{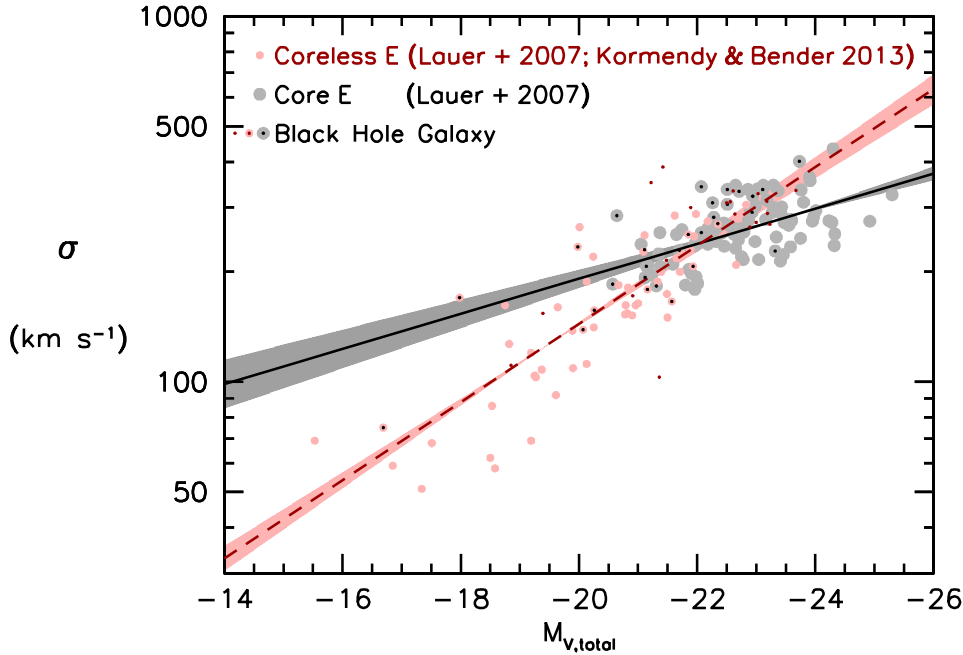} 
\end{center}
\vspace*{-0.3 cm}
\caption{Faber-Jackson (1976) correlations for ellipticals with and without a ``core'', i.{\thinspace}e., 
a break in the density profile near the center from a steep outer profile to a shallow inner cusp.  
Total $V$-band absolute magnitudes and velocity dispersions are from Lauer et al.~(2007) with corrections 
from Kormendy et al.~(2009) and Kormendy \& Bender (2013).  This figure is from the latter paper, 
here labeling galaxies that have BH detections from spatially resolved near-central dynamics.  The lines
are symmetric least-squares fits (Tremaine et al.~2002) to the core galaxies (solid line with gray shading) 
and the coreless galaxies (dashed line with lighter shading of 1-$\sigma$ fit uncertainties).  The kink at 
$\sigma \simeq 250$ km s$^{-1}$ was also emphasized by Lauer et al.~(2007).}
\label{fig1}
\end{figure}

\newpage

\noindent galaxies that are not known to be BH hosts.  However, {\it the reason is not that we looked for BHs and failed.
If that were the case, then Shankar's criticism would be valid.}  Rather, these are (e.{\ts}g., more distant)
galaxies that have not been searched for BHs.  The BH correlations in Figure 3 do not show a kink at $M_K \simeq -25$
corresponding to $M_V \simeq -22$ as would be the case if $M_\bullet$ values were fairly sampled at $M_V > -22$ 
but, at $M_V < -22$, were overestimated by an ``even larger'' factor than ``at least $\sim$3'' as Shankar suggests.

      Note in Figure 2 that coreless ellipticals have the well known $L_V$\ts$\propto$\ts$\sigma^4$
correlation, whereas core ellipticals have an $L_V$\ts$\propto$\ts$\sigma^8$ 
correlation.  These are understood to result, respectively, from ``wet'' major mergers with cold gas dissipation
and central starbursts and dissipationless, ``dry'' mergers in which $\sigma$ grows only slowly with 
luminosity $L_V$.

      Shankar and van den Bosch believe that typical galaxies are more diffuse than BH hosts
because they compare to large samples of galaxies measured by the Sloan Digital Sky Survey.
These must mainly be far away, so spectroscopic apertures sample large radii that include
outward $\sigma$ decreases and{\ts}--{\ts}more importantly\ts--{\ts}galaxy disks.  Also, they do not
make bulge-disk decompositions but rather include galaxy disks and pseudobulges in single measurements of
$r_e$, $\mu_e$, and $\sigma$.  Pseudobulges and disks are fluffier than classical bulges and ellipticals.
They do not participate in BH{\ts}--{\ts}host correlations (Section 5).

\vskip -25pt \null

\section{M{\lower 2pt \hbox{$\bullet$}} -- Host Galaxy Correlations for Classical Bulges and Ellipticals}
 
      Figure 3 shows the correlations of $M_\bullet$ with the $K$-band luminosity and
the velocity dispersion of the host bulge measured outside the BH sphere of influence.   The intrinsic scatter is 
essentially the same for both correlations, respectively, 0.30 and 0.28 dex.  Using

\vfill

\begin{figure}[b]
\vspace*{-2.2 cm}
\begin{center}
\includegraphics{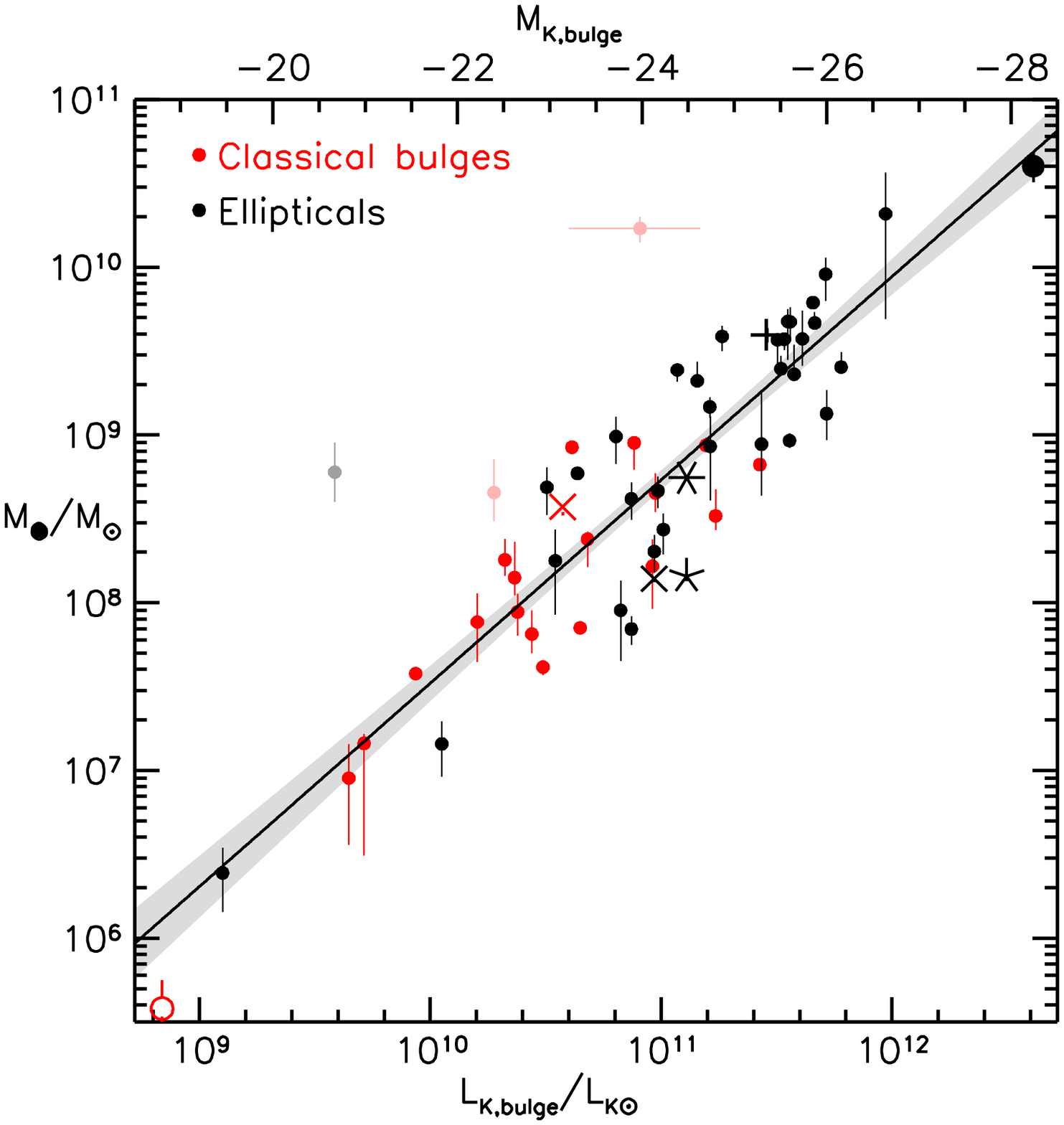}
\includegraphics{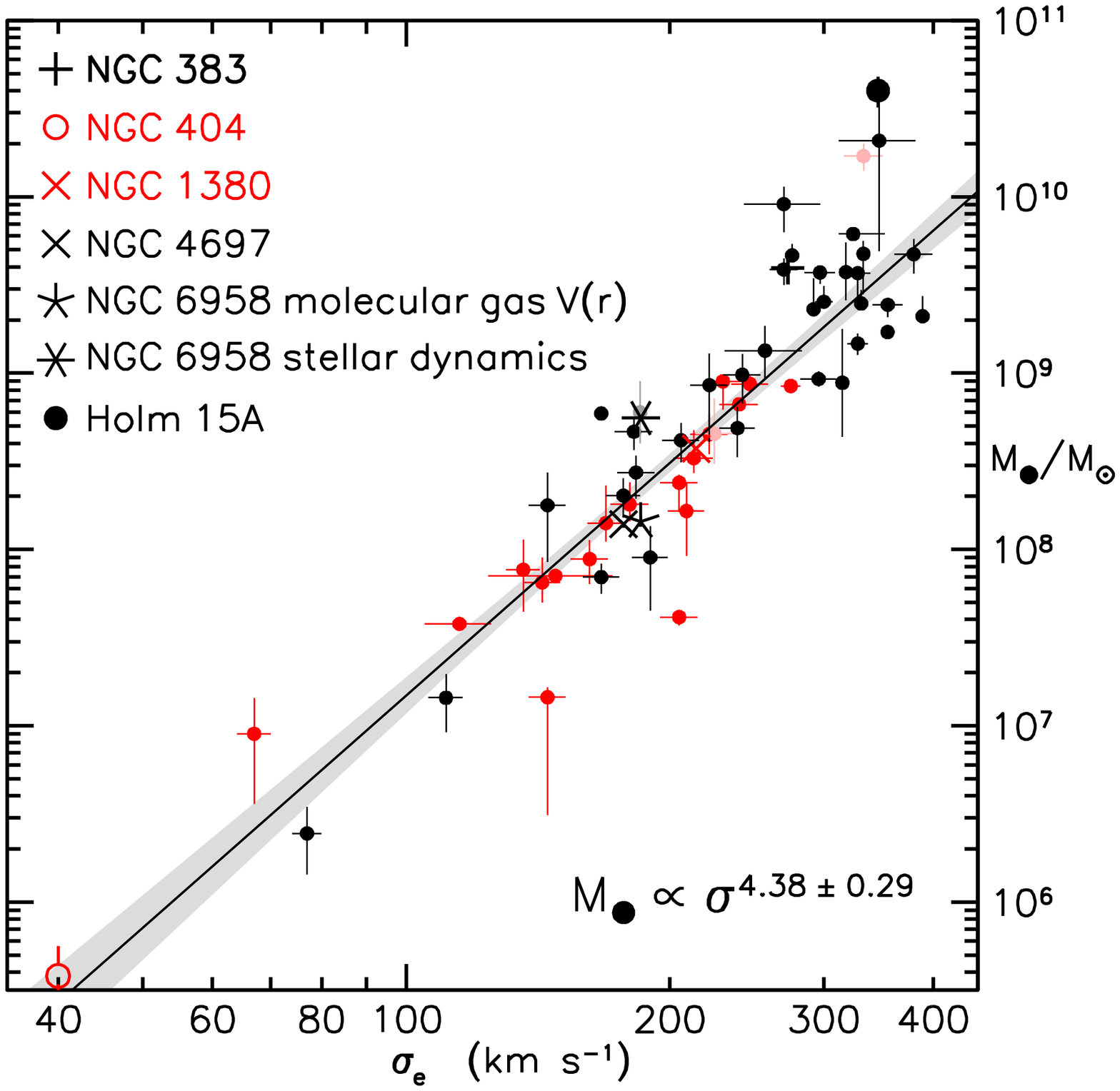}
\end{center}
\vspace*{-0.0 cm}
\caption{Correlations of BH mass with ({\it left}) $K$-band luminosity and
absolute magnitude and ({\it right}) velocity dispersion of the host elliptical galaxy or classical bulge.
This is Figure~17~of Kormendy \& Ho (2013) with six galaxies added (see the key).  Lines and shaded 1-$\sigma$ 
uncertainties are symmetric least-squares fits to the Kormendy \& Ho sample. The fits omit BH monsters that 
have BH mass fractions of $>$\thinspace10\thinspace\% (two lightly shaded~points).  Important: this figure omits 
pseudobulges and major mergers in progress.  Pseudobulges show no tight $M_\bullet$\thinspace--{\thinspace}host 
correlations.  Mergers in progress also do not participate in these correlations: they have undermassive BHs.
Only classical bulges and ellipticals as shown here participate in tight $M_\bullet$ correlations that 
are suggestive of BH\thinspace--{\thinspace}host coevolution. Figures 1 and 2 show that these objects 
fairly sample the fundamental plane correlations. We see no signs that the detected BH masses are biased 
by sample selection.  And the new BH detections mostly added at this meeting ({\it key}) extend and are consistent 
with the $M_\bullet$\thinspace--{\thinspace}host correlations.}
\label{fig1}
\end{figure}

\newpage

\noindent mass-to-light ratios to convert $L_K$ to bulge stellar mass $M_{\rm bulge}$ (Figure 4, bottom left),
\vskip -12pt
$$ 
\quad{{M_\bullet} \over {10^9~M_\odot}} = \biggl(0.49^{+0.06}_{-0.05}\biggr)\ 
                                           \biggl({{M_{\rm bulge}} \over 
                                                  {10^{11}{\ts}M_{\odot}}}\biggr)^{1.17 \pm 0.08};~
                           {\rm intrinsic~scatter} = 0.28~{\rm dex}. \eqno{(1)}
$$
\vskip -5pt
The canonical BH-to-bulge mass ratio, 
$M_\bullet/M_{\rm bulge} = 0.49^{+0.06}_{-0.05}$\ts\% at $M_{\rm bulge} = 10^{11}$\ts$M_\odot$,
is 2--4 times larger than previous values, because (1) we omit pseudobulges; they do not satisfy 
$M_\bullet$ correlations; (2) we omit galaxies with $M_\bullet$ measurements based on ionized gas dynamics that 
do not account for broad emission-line widths; (3) we omit mergers~in progress:~Kormendy \& Ho show that 
these have undermassive black holes, in part because progenitor disks have undermassive BHs.  And (4) many 
BH masses were revised upward when halo DM and more complete orbit sampling were added to dynamical models.

      To further address possible bias in deriving BH{\ts}--{\ts}host correlations, I add
six new BH mass measurements to Figure\ts3, five of them reported at this meeting.  They are identified in the key; 
NGC\ts1380 (Tsukui 2019); 
NGC\ts383, NGC\ts404, and NGC\ts4697 (Bureau 2019), 
NGC\ts6958 (Thater 2019), and 
Holm 15A (Mehrgan et al.~2019).  

      Bureau's measurement in NGC\ts4697 of $M_\bullet = 1.39^{+0.07}_{-0.03} \times 10^8$ $M_\odot$ via an 
ALMA molecular gas rotation curve (scaled to distance 12.54 Mpc as in Kormendy \& Ho 2013) is slightly smaller 
than but consistent with $M_\bullet = 2.02^{+0.51}_{-0.50} \times 10^8$ $M_\odot$ from stellar dynamics.  
The same appears true of NGC\ts6958 (all new $M_\bullet$ measurements reported at this meeting are preliminary).  
If differences persist between ALMA-derived and stellar dynamical $M_\bullet$ measurements, 
this may point to a systematic error in one or both techniques.  If the problem is stellar dynamics, then a
likely culprit is the assumption that $M/L$ is independent of radius.  Allowing stellar $M/L$ to increase 
toward galaxy centers may decrease BH mass estimates by a few tens of percents (McConnell et al.~2013).

      All new galaxies except Holm 15A were measured using molecular gas rotation~curves.
Bureau (2019) emphasizes that they have different systematics from other $M_\bullet$ machinery.
Despite his concerns about biases, they are consistent with the Figure 3 correlations.

      Agreement between stellar dynamics, ionized gas dynamics, megamaser dynamics, and ALMA-based
molecular gas dynamics supports our conclusion (Kormendy \& Ho 2013) that the BH{\ts}--{\ts}host 
bulge correlations are robust.  The discussion of Section 3 further reassures us that biased BH samples
do not significantly affect the derived correlations.  
      
      \underline{Important:} This conclusion applies only to $M_\bullet$ correlations for classical bulges and 
ellipticals.  Section 5 reviews the evidence that $M_\bullet$ shows no such strong correlations with pseudobulges, 
disks, or dark matter halos. BH searches in such objects have frequently yielded only $M_\bullet$ upper limits.  The 
BH masses observed in these components almost certainly are upper envelopes of BH mass distributions that 
extend to lower masses.

      \underline{Returning to astrophysical conclusions:} Figure 3 adds the biggest BH discovered so far in 
the nearby Universe, $M_\bullet = (4.0 \pm 0.8) \times 10^{\rm 10}$ $M_\odot$ in Holm 15A, the giant elliptical
with the largest known core (Mehrgan et al.~2019).  It extends the $M_\bullet$\ts--\ts$L_{K,\rm bulge}$ 
correlation toward larger luminosities and strengthens an important conclusion that has become evident as 
we find more giant BHs.  As concluded by Kormendy \& Ho (2013) and as seen also by McConnell et al. (2011), 
the $M_\bullet$\ts--\ts$\sigma$ correlation ``saturates'' at $\sigma \sim 250$~km~s$^{-1}$.  For velocity 
dispersions higher than this value, $\sigma$ grows slowly or not at all as $M_\bullet$ increases.  Figure\ts2 
indicates why: Core galaxies are remnants of dry mergers in which $\sigma$ grows only slowly while $M_{\rm bulge}$ 
and $M_\bullet$ grow rapidly in successive mergers. If the Faber-Jackson relation has a kink at $M_V \sim -22$ 
and $\sigma \sim 250$ km s$^{-1}$, then $M_\bullet$\ts--\ts$L$~and $M_\bullet$\ts--\ts$\sigma$ cannot both be 
straight lines in Figure\ts3, without a kink.  Holm 15A and other giant ellipticals define a single, straight 
correlation between $\log{M_\bullet}$ and $\log{L}$ or $\log{M_{\rm bulge}}$ (Equation 1).  The more famous 
$M_\bullet$\ts--\ts$\sigma$ correlation has a kink at large $M_\bullet$~and~$\sigma$.  This possibility was
foreseen by Lauer et al.~(2007).

\section{M{\lower 2pt \hbox{$\bullet$}} Does Not Correlate with Pseudobulges, Disks, or Dark Matter}

      Figure 4 repeats the $M_\bullet$ correlations with $K$-band luminosity and velocity dispersion
of the host classical bulge and elliptical and adds the correlation with bulge stellar mass derived using
mass-to-light ratios zeropointed to SAURON dynamical models (Cappellari et al.~2006; Williams et al.~2009).
Pseudobulges are added in light blue shading.  They demonstrate the result (references in Figure 4 key) that 
BH masses do not correlate tightly enough with pseudobulges to be suggestive of BH-host
coevolution.  Since coevolution generally involves feedback that controls star formation, it is worth
emphasizing that pseudobulges are {\it not\/} quenched in their star formation.  Rather, they form stars very
vigorously, at least as vigorously as their associated disks 
(Kormendy \& Kennicutt 2004;
Fisher 2006;
Fisher et al. 2009).

      Recognition of pseudobulges as distinct from elliptical-galaxy-like classical bulges is part of a
picture of the slow, ``secular'' evolution of disk galaxies that complements our picture of 
hierarchical clustering (Kormendy 1993; Kormendy \& Kennicutt 2004; Kormendy 2013 provide reviews).
Prototypical examples occur in the globally oval disk galaxies NGC\ts1068 (Figure\ts4), NGC\ts4151, and
especially NGC\ts4736.  Bars and oval disks transport disk angular momentum outward and rearrange disk
gas into outer rings (NGC\ts1068 shows an example), inner rings that encircle the ends of bars, and gas
that falls toward the center.  There, high gas densities drive strong star formation as described \phantom{00000}

\begin{figure}[b]
\vspace*{-2.2 cm}
\begin{center}
\includegraphics[width=5.3in]{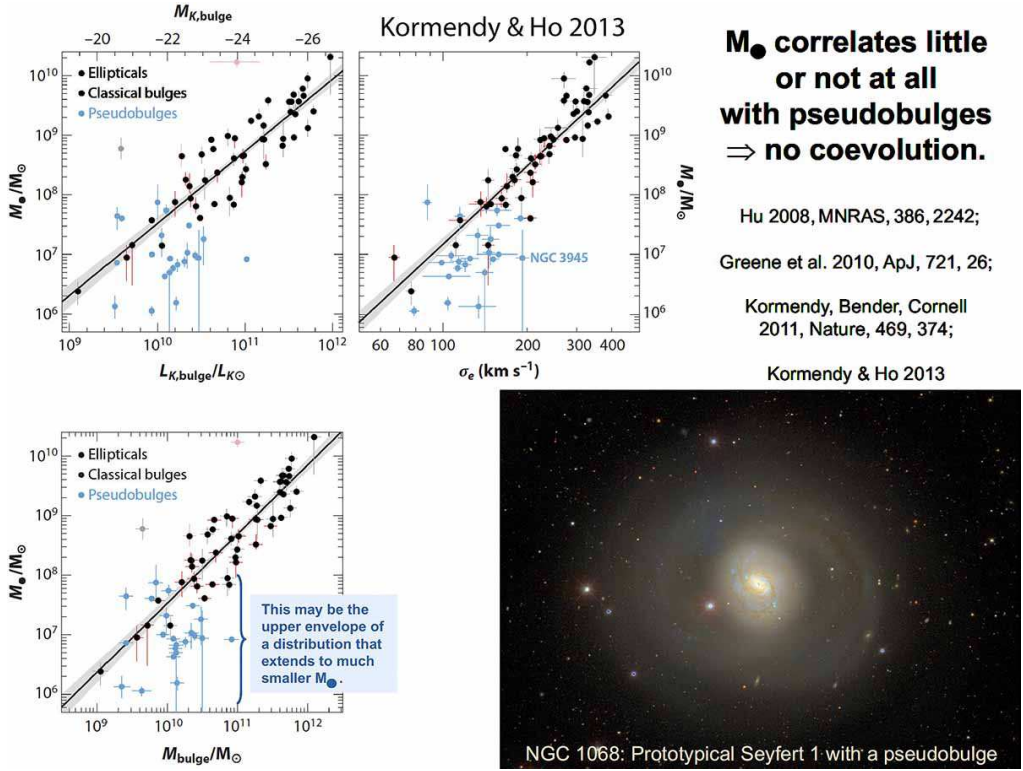} 
\end{center}
\vspace*{-0.15 cm}
\caption{Powerpoint slide showing the correlations (or lack of correlations) of BH mass~$M_\bullet$~with 
({\it left}) $K$-band luminosity and stellar mass and ({\it right}) velocity dispersion of the host elliptical,
classical bulge, or pseudobulge.  Classical bulges and ellipticals participate~in~tight correlations 
with intrinsic scatter $\sim$\ts0.28 dex.  {\it Pseudobulge BHs ({lightly shaded}) show no significant correlation
with their hosts.}  This is Figure~21~of Kormendy \& Ho (2013) adding references and a Sloan Digital Sky Survey 
$gri$ color image of the Seyfert galaxy NGC{\thinspace}1068, a prototypical unbarred but oval (R)SAb galaxy with a 
pseudobulge.  Another such galaxy is the Seyfert NGC{\thinspace}4151.}
\label{fig1}
\end{figure}

\newpage

\noindent by the Schmidt (1959)\ts--{\ts}Kennicutt (1998a,{\ts}b) law.  The result is to make both ring types 
visible in starlight and to build, near the center, a compact stellar component that was grown slowly out of the disk,
not assembled rapidly via galaxy mergers.  These high-density centers were misidentified as classical bulges by early 
morphologists
(e. g., Hubble 1936; 
de Vaucouleurs 1959; 
Sandage 1961) 
and so are called ``fake bulges'' or ``pseudobulges''.  They recognizably remember their disky origin -- 
they are flatter than classical bulges, often as flat as their outer disks; 
they are more rotationally supported, i.{\ts}e., $V/\sigma$ is larger than in classical bulges;
they often show spiral structure and nuclear bars that can only be sustained in flat, cold disks, and (except in S0s)
they show vigorous star formation.  Central components in BH disk galaxies were classified as classical or pseudo 
before the BH correlations were derived.  It is a success of the secular evolution picture that classical and 
pseudo bulges then prove to correlate differently with their BHs.  Many of these BHs are actively accreting AGNs
(e.{\ts}g., NGC\ts1068 and NGC\ts4151).  This is part of the evidence that feedback from AGNs 
does not quench the star formation or otherwise influence the evolution of galaxy disks.

      A further indication that BHs do not affect disk evolution is the observation that BH masses
do not correlate with the $K$-band luminosities and hence the stellar masses of disks.  This is shown in
Figure 5.  Not surprisingly, for disk galaxies with BH discoveries, $M_\bullet$ correlates less well with 
total galaxy luminosity and stellar mass than it does with classical bulge properties.

\begin{figure}[b]
\vspace*{-2.2 cm}
\begin{center}
\includegraphics[width=4.2in]{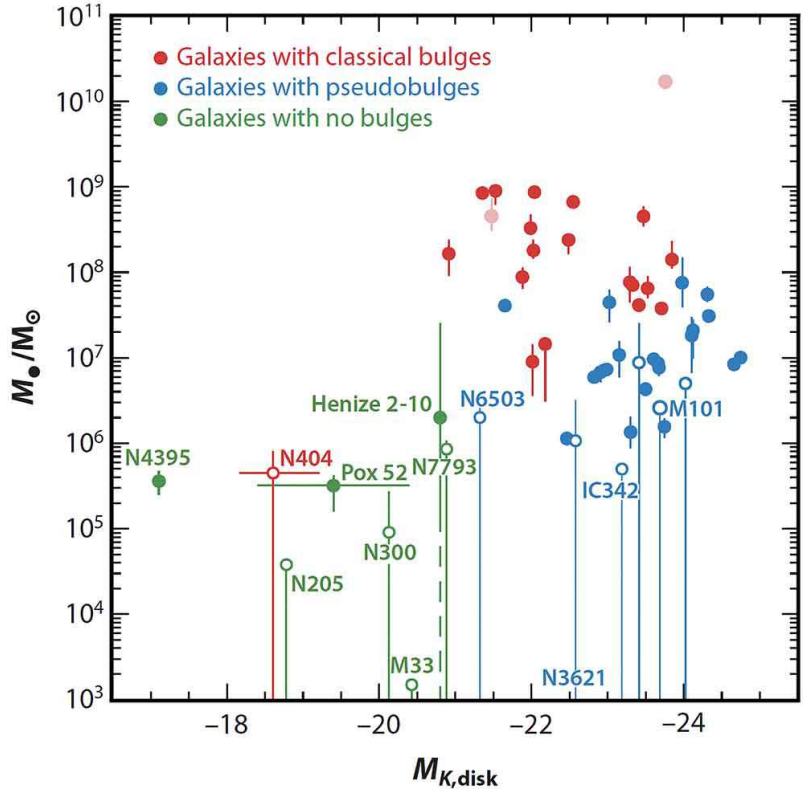} 
\end{center}
\vspace*{-0.15 cm}
\caption{BH mass vs.~$K$-band absolute magnitude of the host disk (Figure 22 of Kormendy \& Ho 2013).  
Solid symbols represent BH detections; open symbols are for BH mass upper limits.
BH masses do not correlate with properties of their host disks.  At M$_{K,{\rm disk}} \simeq -21 \pm 1$,
BH masses range from detections with $M_\bullet \simeq 10^9$\ts$M_\odot$ to the upper limit of 1500 $M_\odot$ in 
M{\thinspace}33 (Gebhardt et al. 2001; Merritt et al. 2001).}
\label{fig1}
\end{figure}

\newpage

      Finally,{\ts\ts}$M_\bullet${\ts\ts}does{\ts}not{\ts}correlate{\ts}with{\ts}DM{\ts}halos{\ts}beyond{\ts}the{\ts}correlation{\ts}implied{\ts}by{\ts}Figure\ts3.
This result is unpopular with galaxy formation theorists, especially numerical modelers who add baryon physics 
to simulations of DM hierarchical clustering.  DM mass $M_{\rm DM}$ is arguably the most fundamental parameter 
associated with a galaxy, and it would have been convenient if it controlled galaxy evolution partly via AGN feedback. 
Unfortunately, this is not the case, as suggested already by the fact that dwarf ellipticals such as M{\ts}32 participate
in tight $M_\bullet$ correlations whereas giant disks such as M{\ts}101 do not.

      Ferrarese (2002) and Baes et~al.~(2003) suggested that $M_\bullet$ {\it does} correlate with $M_{\rm DM}$.  
Their conclusions seem to reflect the conspiracy that baryons and DM are arranged~so $V(r)$ is featureless
even though baryons dominate at small $r$ and DM dominates at large~$r$.  

      Kormendy \& Bender (2011) and Kormendy \& Ho (2013) present seven arguments against the hypothesis that $M_\bullet$ 
correlates fundamentally with $M_{\rm DM}$.  One is illustrated in Figure\ts6.  The correlation of galaxy stellar
mass $M_*$ with DM mass $M_{\rm DM}$ is complicated.  The ratio $M_*/M_{\rm DM}$ is largest at $M_{\rm DM} \equiv M_{\rm crit} = 10^{12}$ $M_\odot$,
the critical mass above which galaxies can gravitationally hold onto large amounts of hot, X-ray-emitting gas.
This keeps baryons increasingly suspended in hot gas\ts--{\ts}not stars\ts--{\ts}in bigger galaxies, accounting
for the decrease in $M_*/M_{\rm DM}$ at larger $M_{\rm DM}$.  The decrease in $M_*/M_{\rm DM}$ at $M_{\rm DM} \ll
10^{12}$ $M_\odot$ is believed to result from more supernova-driven baryon ejection in smaller galaxies.  As a result, 
the correlation of $M_\bullet$ with $M_{\rm DM}$ must be very different at $M_{\rm DM} \ll 10^{12}$ $M_\odot$ and 
at $M_{\rm DM} \gg 10^{12}$ $M_\odot$.  But the correlation of $M_\bullet$ with $M_* \equiv M_{\rm bulge}$
is continuous across $M_{\rm DM} = 10^{12}$ $M_\odot$.  Thus Equation 1 appears to be the fundamental~correlation.

\begin{figure}[b]
\vspace*{-2.2 cm}
\begin{center}
\includegraphics[width=4.9in]{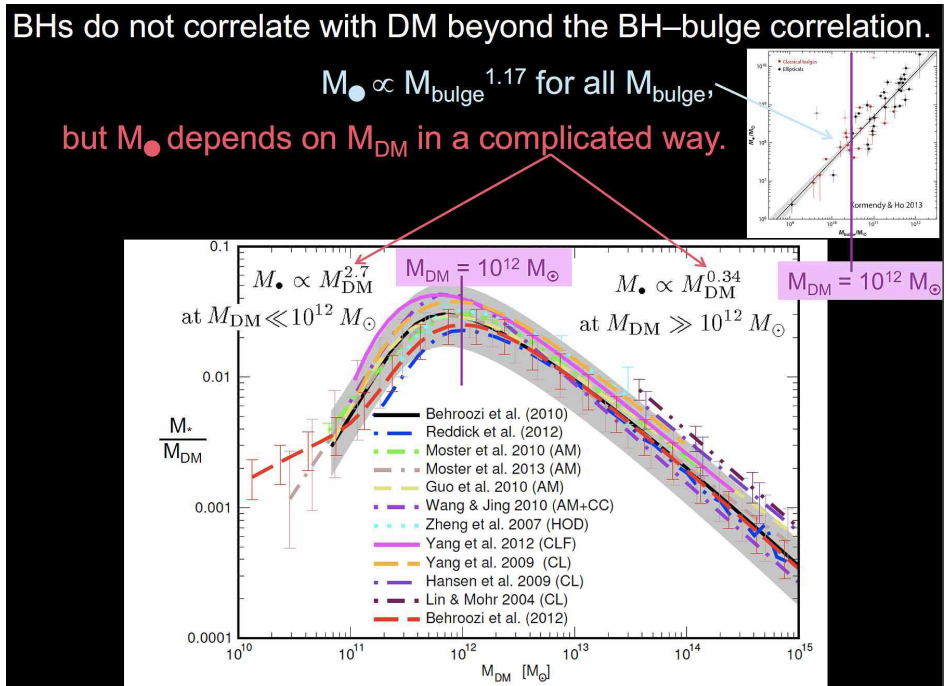} 
\end{center}
\vspace*{-0.2 cm}
\caption{Powerpoint slide illustrating one argument why we conclude (Kormendy \& Bender 2011; 
Kormendy \& Ho 2013) that $M_\bullet$ does not correlate with dark matter (DM) halos beyond
the known correlation with the masses of classical bulges and ellipticals.  The correlation (inset) of
$M_\bullet$ with {\it stellar} masses $M_*$ of bulges and ellipticals is continuous across the
dark matter transition mass $M_{\rm DM} \simeq 10^{12}$ $M_\odot$, but the 
correlation of stellar mass $M_*$ (or $M_*/M_{\rm DM}$ as illustrated here from Behroozi
et al. 2013) changes slope sharply at $M_{\rm DM} \simeq 10^{12}$ $M_\odot$.  Therefore the
correlation of $M_\bullet$ with $M_{\rm DM}$ must be different at $M_{\rm DM} \gg 10^{12}$
$M_\odot$ and at $M_{\rm DM} \ll 10^{12}$~$M_\odot$.  Arrows point to these different correlations.  
The references in the key are not included here.}
\label{fig1}
\end{figure}

\newpage

\section{Four Regimes of BH -- Host Galaxy Coevolution}

\def\gapprox{$_>\atop{^\sim}$} \def\lapprox{$_<\atop{^\sim}$}

      This review concentrates on the demographics of BHs discovered via spatially resolved stellar
or gas dynamics in and near the sphere-of-influence radius of the BH.  BHs in classical bulges and 
ellipticals are typically discovered using spatial resolution \gapprox 4 times the radius of the
BH sphere of influence (Kormendy \& Ho 2013, Figure 1).  This was true even for the BHs discovered 
using ground-based spectroscopy before HST became available.  Most of these BHs could still 
have been discovered if they were several times less massive.  But they weren't. The conclusion that 
$M_\bullet$ correlates tightly with the luminosity, stellar mass, and velocity dispersion of host classical 
bulges and ellipticals is robust.
 
      On the other hand, $M_\bullet$ is found not to correlate with any other structural component in
galaxies -- not pseudobulges, not disks, and not dark matter halos.  This, together with other evidence 
reviewed in Kormendy \& Ho (2013), allows us to refine our picture of when BHs do and when they do not coevolve 
with their hosts.  We suggest that there are four distinct regimes of coevolution. Quoting from our review: 

``(1) Local, secular, episodic, and stochastic feeding of small BHs in largely bulgeless galaxies
involves too little energy to result~in~coevolution.''  

``(2) Global feeding in major, wet galaxy mergers rapidly grows giant BHs in 
short-duration, quasar-like events whose energy feedback does affect galaxy evolution.  
The resulting hosts are classical bulges and coreless-rotating-disky ellipticals.''

``(3) After these AGN phases and at the highest galaxy masses, maintenance-mode BH feedback into X-ray-emitting 
gas has the primarily negative effect of helping to keep baryons locked up in hot gas and thereby keeping 
galaxy formation from going to completion.  This happens in giant, core-nonrotating-boxy ellipticals.  
Their properties, including tight correlations between $M_\bullet$ and core parameters [Kormendy \& 
Bender 2009], support the conclusion that core ellipticals form by dissipationless major mergers.  
They inherit coevolution effects from smaller progenitor galaxies.'' 

``(4) Independent of any feedback physics, in BH growth modes 2 and 3, the averaging that results 
from successive mergers plays a major role in decreasing the scatter in $M_\bullet$ correlations 
from the large values observed in bulgeless and pseudobulge galaxies to the small values observed in giant 
elliptical galaxies'' (Peng 2007; Gaskell 2010; Hirschmann et al.~2010; Jahnke \& Macci\`o 2011).  
It is no accident that pseudobulge BH masses range from the largest
$M_\bullet$ observed in ellipticals of similar mass down to much smaller masses.  Mergers convert
pseudobulges into classical bulges and ellipticals, adding new stars in starbursts, merging the 
progenitor BHs, and in general growing the resulting BHs further by gas accretion.

\vskip -27pt

\centerline{\null}

\section{Conclusions}

Figure 7 lists astrophysical conclusions of this paper, echoing Kormendy \& Ho (2013).

I also emphasize the practical conclusion of this paper:

Selection effects do not invalidate the $M_\bullet$ correlations shown here
and in Kormendy \& Ho (2013).  Only classical bulges and elliptical galaxies
participate in tight correlations of $M_\bullet$ with the stellar mass and velocity dispersion 
of the host (intrinsic~scatter~$\simeq$~0.28~dex).  Classical bulges and 
ellipticals are well sampled over the complete range of their masses; objects with
BH detections sample essentially completely the tight fundamental plane correlations of their hosts.
In contrast, BH searches frequently fail for pseudobulges and disks: the BHs that we detect in these objects
are very likely to be the high-$M_\bullet$ part of a scatter that extends to
lower $M_\bullet$ masses.  Thus our picture of disk secular evolution (Kormendy 1993; 
Kormendy \& Kennicutt 2004; Kormendy 2013) is an indispensible part of our understanding
of how BHs do and do not coevolve with their host galaxies.

\newpage

\centerline{\null} \vskip 3.55truein

\includegraphics{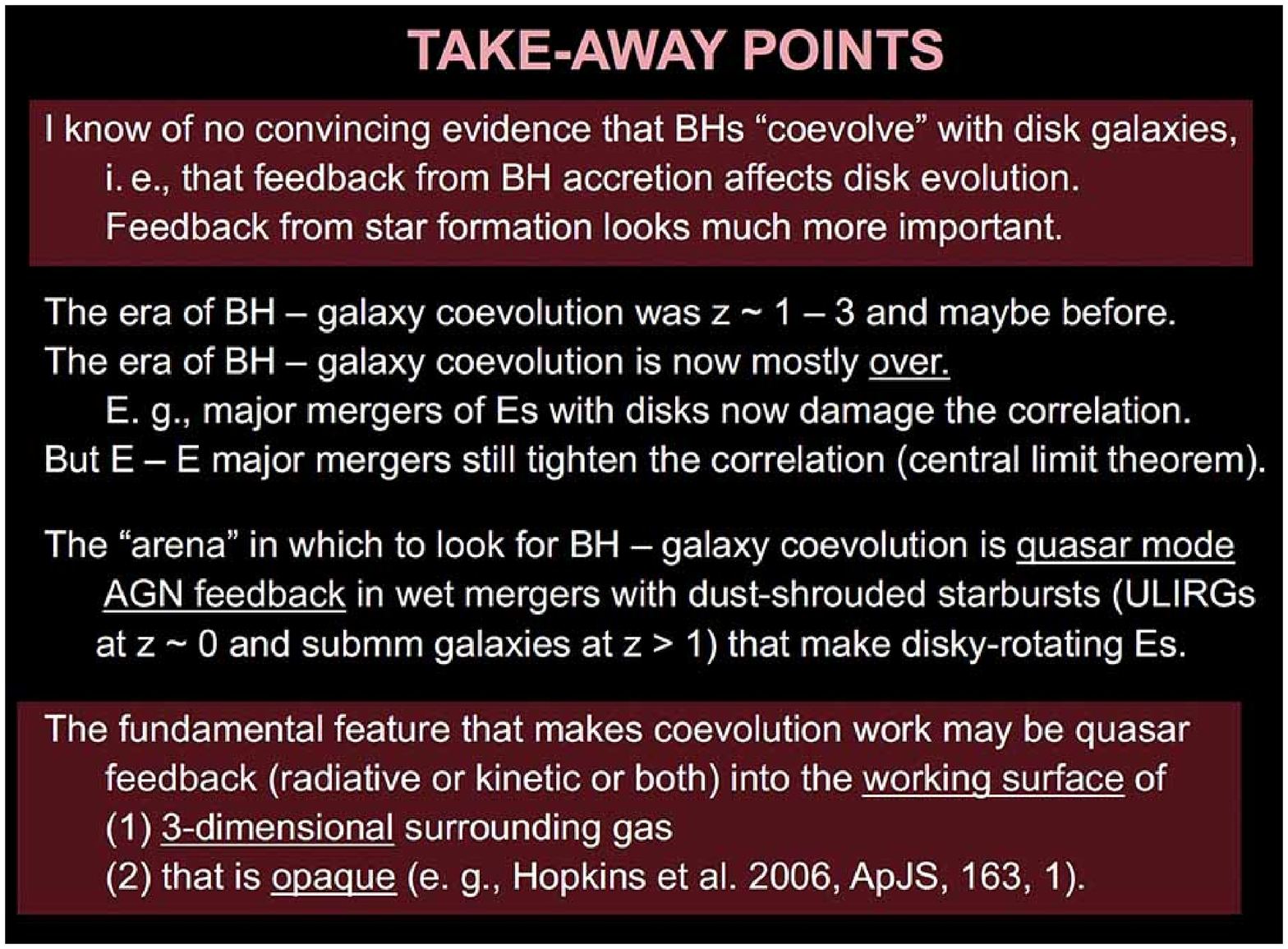}

\centerline{{\bf Figure 7.}~Powerpoint slide of astrophysical conclusions of this paper.}

\vskip 10pt
\noindent {\bf Acknowledgments}
\vskip 5pt

      It is a pleasure to thank Ralf Bender and Luis Ho for many years of very enjoyable and fruitful 
collaboration.  I thank Scott Tremaine for helpful comments on this work.  And I am grateful to Ralf Bender 
and the Max Planck Institute for Extraterrestrial Physics and to Juntai Shen, the Shanghai Astronomical Observatory, 
and Shanghai Jiao Tong University for my financial support to attend IAU Symposium 353.

\vskip 10pt
\noindent {\bf References}
\vskip 5pt

\frenchspacing
\def\nhi{\noindent  \hangindent=1.0truecm}

\nhi Baes, M., Buyle, P., Hau, G. K. T., \& Dejonghe, H. 2003, MNRAS, 341, L44

\nhi Baggett, W. E., Baggett, S. M., \& Anderson, K. S. J. 1998, AJ, 116, 1626

\nhi Behroozi, P. S., Wechsler, R. H., \& Conroy, C. 2013, ApJ, 770, 57

\nhi Bender, R., Burstein, D., \& Faber, S. M. 1992, ApJ, 399, 462

\nhi Bureau, M. 2019, paper presented at IAU Symposium 353, Galactic Dynamics in the Era of Large Surveys,
     posted at {\tt https://pan.cstcloud.cn/s/OaHRZGsQTgs}

\nhi Cappellari, M., Bacon, R., Bureau, M., et al. 2006, MNRAS, 366, 1126

\nhi de Vaucouleurs, G. 1959, Handbuch der Physik, 53, 275

\nhi Event Horizon Telescope Collaboration 2019, ApJ, 875, L6

\nhi Faber, S. M., \& Jackson, R. E. 1976, ApJ, 204, 668

\nhi Ferrarese, L. 2002, ApJ, 578, 90

\nhi Fisher, D. B. 2006, ApJ, 642, L17

\nhi Fisher, D. B., \& Drory, N. 2008, AJ, 136, 773

\nhi Fisher, D. B., Drory, N., \& Fabricius, M. H. 2009, ApJ, 697, 630

\nhi Gaskell, C. M. 2010, in The First Stars and Galaxies: Challenges for the Next Decade, 
     ed. D. J. Whalen et al. (Melville, NY: AIP), 261

\nhi Gavazzi, G., Franzetti, P., Scodeggio, M., Boselli, A., \& Pierini, D. 2000, A\&A, 361, 863

\nhi Gebhardt, K., Adams, J., Richstone, D., et al. 2011, ApJ, 729, 119

\nhi Gebhardt, K., Lauer, T. R., Kormendy, J., et al. 2001, AJ, 122, 2469

\nhi Gebhardt, K., \& Thomas, J. 2009, ApJ, 700, 1690

\nhi Greene, J. E., Peng, C. Y., Kim, M., et al. 2010, ApJ, 721, 26

\nhi Hirschmann, M., Khochfar, S., Burkert, A., et al. 2010, MNRAS, 407, 1016

\nhi Hopkins, P. F., Hernquist, L., Cox, T. J., et al. 2006, ApJS, 163, 1

\nhi Hu, J. 2008, MNRAS, 386, 2242

\nhi Hubble, E. P. 1936, The Realm of the Nebulae (New Haven: Yale University Press)

\nhi Jahnke, K., \& Macci\`o, A. V. 2011, ApJ, 734, 92

\nhi J\o rgensen, I., Franx, M., \& Kj\ae rgaard, P. 1996, MNRAS, 280, 167

\nhi Kennicutt, R. C. 1998a, ApJ, 498, 541

\nhi Kennicutt, R. C. 1998b, ARA\&A, 36, 189

\nhi Kormendy, J. 1987, ApJ, 218, 333

\nhi Kormendy, J. 1993, in IAU Symposium 153, Galactic Bulges, ed. H. Dejonghe \&
     H. J. Habing (Dordrecht: Kluwer), 209

\nhi Kormendy, J. 2013, in XXIII Canary Islands Winter School of Astrophysics,
     Secular Evolution of Galaxies, ed. J. Falc\'on-Barroso \& J. H. Knapen
     (Cambridge: Cambridge University Press), 1 (arXiv:1311.2609)

\nhi Kormendy, J., \& Bender, R. 2009, ApJ, 691, L142

\nhi Kormendy, J., \& Bender, R. 2011, Nature, 469, 377

\nhi Kormendy, J., \& Bender, R. 2012, ApJS, 198, 2

\nhi Kormendy, J., \& Bender, R. 2013, ApJ, 769, L5

\nhi Kormendy, J., Bender, R., \& Cornell, M. E. 2011, Nature, 469, 374

\nhi Kormendy, J., Fisher, D. B., Cornell, M. E., \& Bender, R. 2009, ApJS, 182, 216

\nhi Kormendy, J., \& Ho, L. C. 2013, ARA\&A, 51, 511

\nhi Kormendy, J., \& Kennicutt, R. C. 2004, ARA\&A, 42, 603

\nhi Lauer, T. R., Faber, S. M., Richstone, D., et al. 2007, ApJ, 662, 808

\nhi Macchetto, F., Marconi, A., Axon, D. J., et al. 1997, ApJ, 489, 579

\nhi McConnell, N. J., Chen, S.-F. S., Ma, C.-P., et al. 2013, ApJ, 768, L21

\nhi McConnell, N. J., Ma, C.-P., Gebhardt, K., et al. 2011, Nature, 480, 215

\nhi Mehrgan, K., Thomas. J., Saglia, R., et al.~2019, submitted to ApJ (arXiv:1907.10608)

\nhi Merritt, D., Ferrarese, L., \& Joseph, C. L. 2001, Science, 293, 1116

\nhi Peng, C. Y. 2007, ApJ, 671, 1098

\nhi Rusli, S. P., Thomas, J., Saglia, R. P., et al. 2013, AJ, 146, 45

\nhi Sandage, A. 1961, The Hubble Atlas of Galaxies (Washington:~Carnegie Institution of Washington)

\nhi Schmidt, M. 1959, ApJ, 129, 243

\nhi Schulze, A., \& Gebhardt, K. 2011, ApJ, 729, 21

\nhi Schwarzschild, M. 1979, ApJ, 232, 236

\nhi Shankar, F., Bernardi, M., Sheth, R. K. et al. 2016, MNRAS, 460, 3119

\nhi Thater, S. 2019, paper presented at IAU Symposium 353, Galactic Dynamics in the Era of Large Surveys,
     posted at {\tt https://pan.cstcloud.cn/s/OaHRZGsQTgs}

\nhi Tremaine, S., Gebhardt, K., Bender, R., et al. 2002, ApJ, 574, 740

\nhi Tsukui, T. 2019, paper preented at IAU Symposium 353, Galactic Dynamics in the Era of Large Surveys,
     posted at {\tt https://pan.cstcloud.cn/s/OaHRZGsQTgs}

\nhi van den Bosch, R. C. E. 2016, ApJ, 831, 134

\nhi van den Bosch, R. C. E., Gebhardt, K., G\"ultekin, K., Y\i ld\i r\i m, A., 
     \& Walsh, J. E. 2015, ApJS, 218, 10

\nhi Williams, M. J., Bureau, M., \& Cappellari, M. 2009, MNRAS, 400, 1665

\end{document}